\documentclass[superscriptaddress,10pt,aps,prl,twocolumn,longbibliography, notitlepage]{revtex4-1}
\usepackage{float}
\floatstyle{boxed}
\usepackage{array}
\usepackage{graphicx}
\usepackage{amsbsy}
\usepackage[utf8x]{inputenc}
\usepackage{epstopdf}
\usepackage{amsmath,amssymb,amsfonts,amsthm}
\usepackage{indentfirst}
\usepackage{soul}
\usepackage[T1]{fontenc}
\usepackage[dvipsnames]{xcolor}
\usepackage{url}
\usepackage[colorlinks]{hyperref}

\hypersetup{
    plainpages=true,
    breaklinks=true,
    hypertexnames=false,
    pageanchor=true,
    colorlinks=true,
    linkcolor={blue},
    citecolor={red},
    urlcolor={blue},
    anchorcolor={black}
}

\newcommand{\figref}[1]{\mbox{Fig.~\ref{#1}}}

\renewcommand{\eqref}[1]{\mbox{Eq.~(\ref{#1})}}

\newcommand{\be}{\begin{equation}}
\newcommand{\ee}{\end{equation}}
\newcommand{\bea}{\begin{eqnarray}}
\newcommand{\eea}{\end{eqnarray}}

\begin{document}

\title{Restoring Adiabatic State Transfer in Time-Modulated Non-Hermitian Systems}
\author{Ievgen I. Arkhipov}
\affiliation{Joint Laboratory of
Optics of Palack\'y University and Institute of Physics of CAS,
Faculty of Science, Palack\'y University, 17. listopadu 12, 771 46
Olomouc, Czech Republic}

\author{Fabrizio Minganti}
\affiliation{ Institute of
Physics, Ecole Polytechnique F\'ed\'erale de Lausanne (EPFL),
CH-1015 Lausanne, Switzerland} \affiliation{Center for Quantum
Science and Engineering, Ecole Polytechnique Fédérale de
Lausanne (EPFL), CH-1015 Lausanne, Switzerland}

\author{Adam Miranowicz}
\affiliation{Theoretical
Quantum Physics Laboratory,  Cluster for Pioneering Research,
RIKEN, Wako-shi, Saitama 351-0198, Japan} \affiliation{Quantum
Information Physics Theory Research Team, Quantum Computing
Center, RIKEN, Wakoshi, Saitama, 351-0198, Japan}
\affiliation{Institute of Spintronics and Quantum Information,
Faculty of Physics, Adam Mickiewicz University, 61-614 Pozna\'n,
Poland}

\author{\c{S}ahin K. \"Ozdemir}
\affiliation{Department of Engineering
Science and Mechanics, and Materials Research Institute (MRI), The
Pennsylvania State University, University Park, Pennsylvania
16802, USA}

\author{Franco Nori}
\affiliation{Theoretical Quantum Physics
Laboratory,  Cluster for Pioneering Research, RIKEN,  Wako-shi,
Saitama 351-0198, Japan} \affiliation{Quantum Information Physics
Theory Research Team, Quantum Computing Center, RIKEN, Wakoshi,
Saitama, 351-0198, Japan} \affiliation{Physics Department, The
University of Michigan, Ann Arbor, Michigan 48109-1040, USA}

\begin{abstract}

Non-Hermitian systems have attracted much interest in recent decades, driven partly by the existence of exotic spectral singularities, known as exceptional points (EPs), where the dimensionality of the system evolution operator is reduced.
Among various intriguing applications,  the discovery of EPs has suggested the potential for implementing a symmetric mode switch, when encircling them in a system parameter space.
However, subsequent theoretical and experimental works have revealed that {\it dynamical} encirclement of EPs invariably results in asymmetric mode conversion; namely, the mode switching depends only on the winding direction but not on the initial state. This chirality arises from the failure of adiabaticity due to the complex spectrum of non-Hermitian systems. Although the chirality revealed has undoubtedly made a significant impact in the field, a realization of the originally sought symmetric adiabatic passage in non-Hermitian systems with EPs  has since been elusive.    
In this work, we bridge this gap and theoretically demonstrate that adiabaticity, and therefore a symmetric state transfer, is achievable when dynamically winding around an EP. This becomes feasible by specifically choosing a trajectory in the system parameter space along which the corresponding evolution operator attains a real spectrum.
Our findings, thus, offer a promise for advancing various wave manipulation protocols in both quantum and classical domains. 
\end{abstract}

\date{\today}

\maketitle

\paragraph{Introduction.---}
Non-Hermitian, i.e., non-conservative or open, systems are characterized by a complex-valued spectrum. Due to this complexity, such systems can exhibit spectral singularities, known as exceptional points (EPs), where both eigenvalues and eigenvectors of a corresponding evolution operator coalesce~\cite{KatoBOOK}. 
The existence of EPs introduces a variety of rich and intriguing phenomena not encountered in conservative, i.e., Hermitian, systems~\cite{Ozdemir2019,Ashida2020,Bergholtz2021}.  

In the quantum realm, the non-Hermitian evolution operator can be represented by either a non-Hermitian Hamiltonian (NHH) or a Liouvillian, depending on whether the portrayal of the system dynamics excludes or includes the effects of quantum jumps, respectively~\cite{BreuerBookOpen}. Consequently, in the description of non-conservative (semi)classical systems or open quantum systems upon postselection, exclusive reliance on the NHH formalism is usually sufficient~\cite{Minganti2020}.

Since the discovery of EPs, it has been anticipated that EPs can be potentially exploited for adiabatic state transfer, due to the Riemann topology of the system spectrum induced by these singularities~\cite{Latinne1995,Lefebvre2009,Atabek2011,Heiss2000,Cartarius2007}. That is, by winding around an EP, one can symmetrically switch between system eigenstates thanks to the the presence of a branch cut between two energy Riemann surfaces~\cite{Guria2024}. Adiabatic evolution implies that this state transfer only depends on an initial state, not on a winding direction. This observation was also confirmed experimentally when realizing {\it stationary}, i.e., time-independent NHHs~\cite{Dembowski2003,Dietz2011,Gao2015,Ding2016,Ergoktas2022}.  

However, subsequent theoretical~\cite{Uzdin2011,Berry2011,Graefe2013} and later experimental~\cite{Doppler_2016,Hassan2017,Xu2016,Yoon2018,Zhang2018_encirc,Zhang2019_encirc2,Feng2022,Tang2023} works have demonstrated that for time-dependent NHHs the adiabaticity assumption breaks down due to the imaginary part of the system spectrum. Namely, when one dynamically encircles an EP, the system inevitably experiences non-adiabatic transitions (NATs), which lead to  a state-flip asymmetry. This appearing chirality ensures that only the orbiting direction determines the final state. Though a recent work~\cite{arkhipov2023b} showed that the state flip symmetry can be recovered in dissipative systems by exploiting the spectral topology of hybrid diabolic-exceptional points~\cite{arkhipov2023c}, but only through non-adiabatic transformations in \textit{multimode} systems.

While the chiral mode behavior revealed in time-modulated non-Hermitian systems has, undoubtedly, led to important advancements in the field, still, achieving the originally sought symmetric adiabatic passage in such systems has remained elusive.
In this work, we bridge this gap and demonstrate that one can restore adiabatic symmetric state transfer in open systems while dynamically orbiting around an EP. This becomes feasible thanks to a specific choice of the encircling trajectory in a system parameter space. 
The  protocol proposed here relies on a proper mapping of the system parameter space of a given NHH onto a certain submanifold, where the NHH becomes pseudo-Hermitian, i.e., a Hamiltonian with real eigenvalues.
Compared to systems with NATs, which are usually associated with instabilities, our protocol exploits the system real spectrum and therefore can provide greater  system control and robustness. The latter property is especially crucial in the quantum domain.
We illustrate our findings with the simplest example of a dissipative two-level system. Our results thus hold promise for advancing light manipulation protocols in both quantum and classical domains.

\begin{figure}[t!]
    \includegraphics[width=0.3\textwidth]{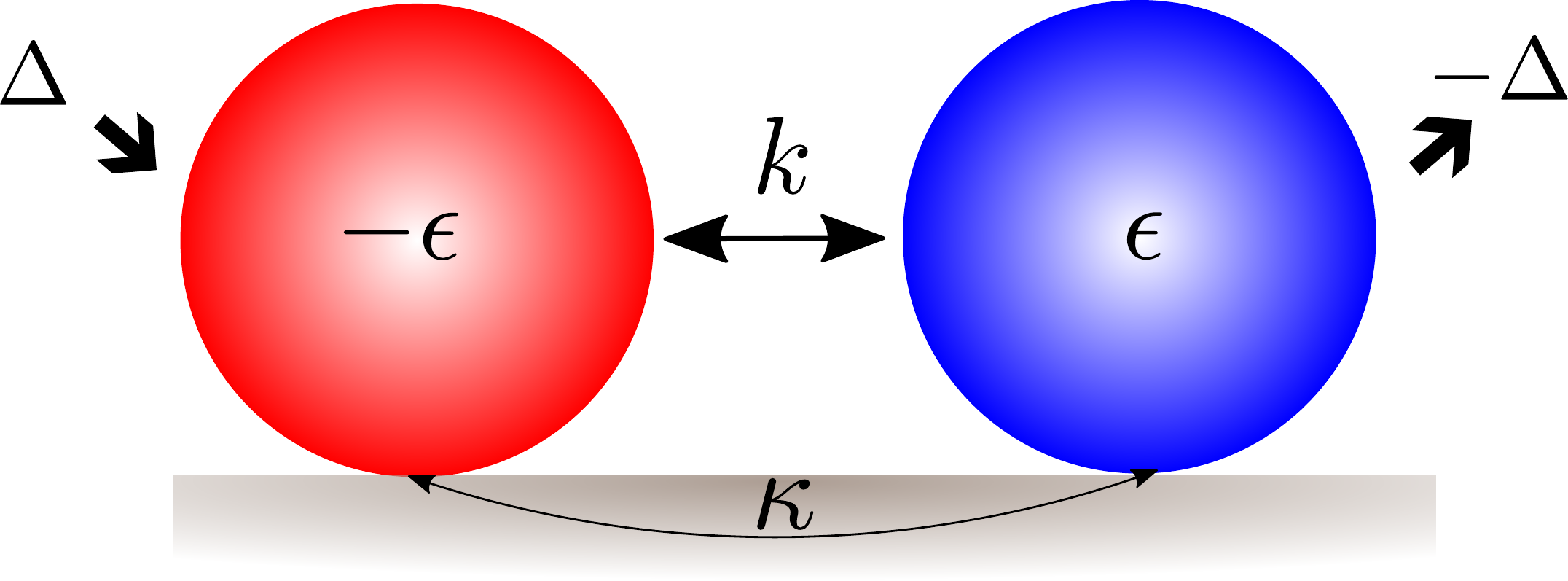}
    \caption{Schematic representation of an open system described by a non-Hermitian Hamiltonian $H$ in \eqref{H_in}. The system consists of two cavities, detuned in frequency $\pm\epsilon$, and coupled coherently with interaction strength $k$, and dissipatively with strength $\kappa$. Both cavities can be  amplified with gain rate $\Delta$ or experience losses with rate $-\Delta$.}
    \label{fig1}
\end{figure}

{\it{Model.---}}
We consider a two-level NHH
\begin{eqnarray}\label{H_in}  
    H = \begin{pmatrix}
        -i\Delta+\epsilon & k+i\kappa \\
        k+i\kappa & i\Delta-\epsilon
    \end{pmatrix},
\end{eqnarray}
describing either a classical two-level system  or a quantum one, subjected to postselection in some global decaying reference frame~\cite{Minganti2020}.
As shown in Fig.~\ref{fig1}, a possible realization of such a system is two coupled dissipative cavities (in the mode representation), where $\Delta$ ($-\Delta$) denotes the resonator gain (loss) rate and $\epsilon$ is the frequency detuning of the resonators. The parameters $k$ and $\kappa$ account for coherent and incoherent, i.e., dissipative, mode coupling strengths, respectively. All parameters are real numbers.
This NHH determines the state evolution via the Schr\"odinger equation
   $ i\partial_t|{\psi}(t)\rangle=H|\psi(t)\rangle$.

The Hamiltonian in \eqref{H_in} has complex eigenvalues $E_{\pm} = \pm \sqrt{[-\Delta +k-i (\epsilon -\kappa )] [\Delta +k+i
(\kappa +\epsilon )]}$ and admits EPs in its parameter space defined by equations $|\Delta_{\rm EP}| = |k|$ and $|\epsilon_{\rm EP}| = |\kappa|$. 
When $\kappa=0$, the NHH in \eqref{H_in} reduces to the paradigmatic classical two-mode model used for the demonstration of the chiral mode behaviour~\cite{Doppler_2016}.
Indeed, while encircling the EP, the imaginary part of the eigenenergies $E_{\pm}$ plays a fundamental role in determining which state `survives' at the end of the winding protocol, due to induced NATs, resulting thus in a chiral state transfer. Evidently, the same conclusion holds true when dynamically encircling other EPs in the system's complex energy space.

{\it{Mapping a non-Hermitian Hamiltonian onto a pseudo-Hermitian one.---}} 
In light of the emergent chiral mode behaviour in non-Hermitian systems with EPs, we are motivated by the following question: Can one dynamically encircle an EP without inducing NATs during the system evolution? In other words, can one restore adiabatic state transfer in a time-modulated non-Hermitian system? Below we show that the answer is affirmative. This becomes possible provided that, along the orbiting trajectory in a system parameter space, a given NHH acquires a pseudo-Hermitian form, i.e., it attains a real spectrum. 

This pseudo-Hermitian transformation of the NHH in \eqref{H_in} can be achieved, in particular, with the help of a certain function $f\!\!\!: \vec{r} = (x,y)\to (\Delta,\epsilon,k,\kappa)$, which maps a two-dimensional (2D) real space $(x,y)$, called a chart, onto a manifold in the $4$D parameter space of the NHH. The minimal dimension of the chart is 2D since for finding an EP one needs only two parameters~\cite{Berry2004}.  The sought  manifold can be parameterized as follows:
\begin{eqnarray}\label{sys_param}
    &\Delta = \alpha\sinh\phi_i\sin\phi_r, \quad \epsilon =\alpha\cosh\phi_i\cos\phi_r,& \nonumber \\
    &k = \alpha\cosh\phi_i\sin\phi_r, \quad \kappa =\alpha\sinh\phi_i\cos\phi_r,&
\end{eqnarray}
where $\phi=\phi_r+i\phi_i=\arctan(x+iy)^{-1}\in{\mathbb C}$, and $\alpha=x\sinh\phi_i/\sin\phi_r\in{\mathbb R}$.
This manifold describes a 4D hyperboloid via
$
    \epsilon^2+k^2-\Delta^2-\kappa^2=\alpha^2.
$
The Hamiltonian in \eqref{H_in} becomes:
\begin{eqnarray}\label{H}
    H = \alpha\begin{pmatrix}
        \cos\phi & \sin\phi \\
        \sin\phi & -\cos\phi
    \end{pmatrix}, \quad E_{1,2}=\mp\alpha.
\end{eqnarray}
The embedding in \eqref{sys_param}, i.e., an injective continuous map, indeed guarantees that the NHH is a pseudo-Hermitian ($H^{\dagger}\neq H$) with  real eigenvalues everywhere on the whole  $(x,y)$-plane, and, therefore, on the corresponding $4$D hyperboloid in the system parameter space.

The corresponding right eigenvectors of $H$ are
\begin{equation}\label{psi}
   |\psi_1\rangle\equiv
        \left[-\sin\frac{\phi}{2},
        \cos\frac{\phi}{2}\right]^T, \quad
    |\psi_2\rangle\equiv
        \left[\cos\dfrac{\phi}{2},
        \sin\dfrac{\phi}{2}\right]^T,
\end{equation}
where $T$ stands for transpose. Together with the left eigenvectors~\footnote{The left eigenvectors $|\eta_k\rangle$ of the NHH $\hat H$ are defined via the equation $\hat H^{\dagger}|\eta_k\rangle=E_k^{*}|\eta_k\rangle$.}
\begin{equation}\label{psi_left}
 |\eta_1\rangle\equiv
        \left[-\sin\dfrac{\phi^*}{2},
        \cos\dfrac{\phi^*}{2}\right]^T, \quad
    |\eta_2\rangle\equiv
        \left[\cos\dfrac{\phi^*}{2},
        \sin\dfrac{\phi^*}{2}\right]^T,
\end{equation}
they form the biorthogonal basis, i.e., $\langle\eta_j|\psi_k\rangle=\delta_{jk}$. 
The introduction of the left, i.e., dual vector space is necessary since the right eigenvectors alone are non-orthogonal~\footnote{
The non-orthogonality of the right eigenvectors can be interpreted as the `non-flatness' of the NHH Hilbert space. 
Such a curved space must be endowed with a metric tensor $g$, and the left eigenvectors allow to determine one, namely, $g\equiv\sum |\eta_i\rangle\langle\eta_i|$~\cite{MOSTAFAZADEH_2010,Ju2019}. The metric $g$ facilitates the proper definition of the inner product on a non-flat space between two arbitrary (right) vectors $|\psi_{j,k}\rangle$ as $\langle \psi_i|\psi_j\rangle_{{\rm flat}}\to\langle \psi_i|g|\psi_j\rangle_{{\rm curved}}$, which ensures that the vector's norm and, therefore, a state probability are preserved in time~\cite{Ju2019}.}.

\begin{figure}
    \includegraphics[width=0.45\textwidth]{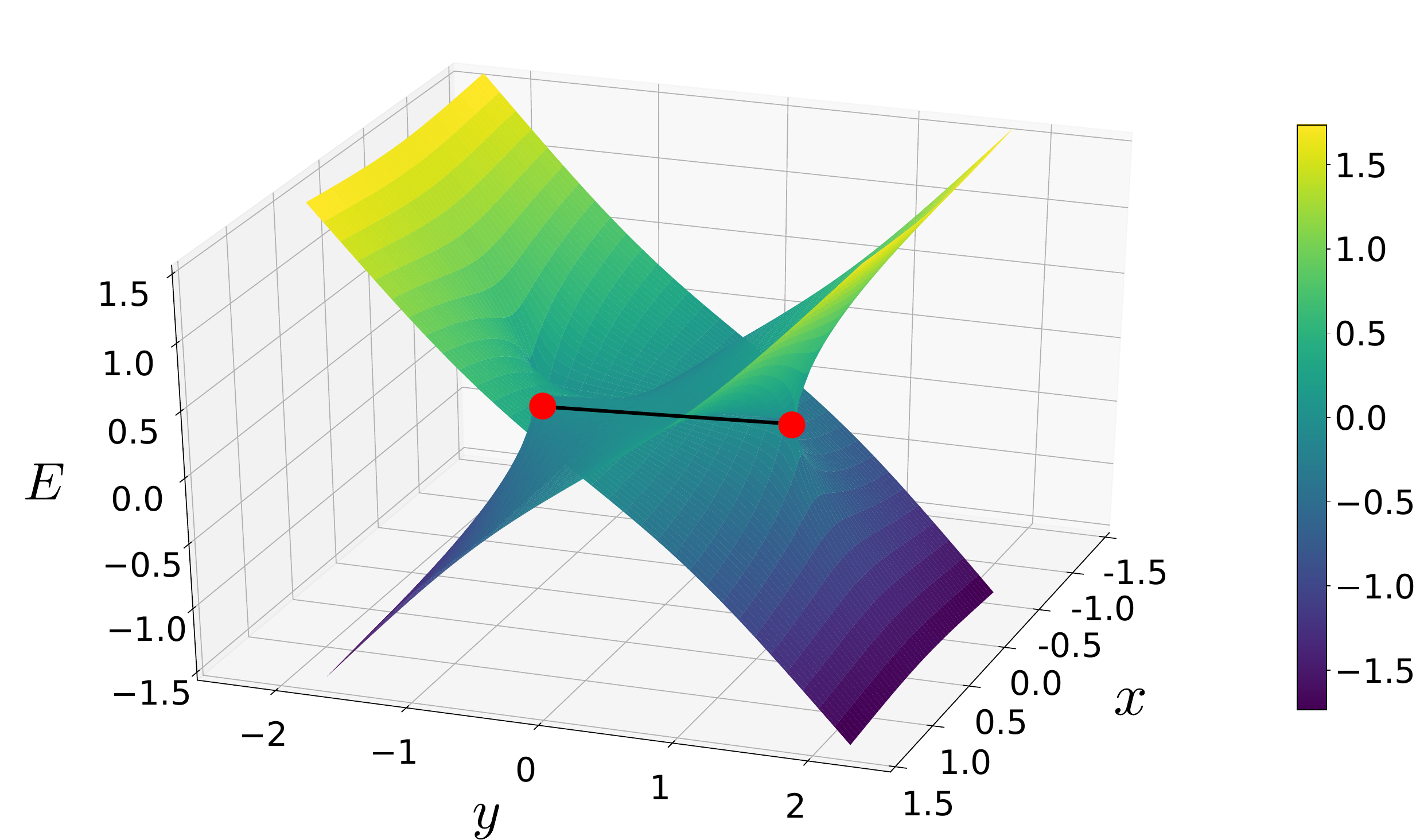}
    \caption{Eigenenergy spectrum $E$ of the pseudo-Hermitian Hamiltonian $H$ given in \eqref{H}.
    The spectrum consists of energy Riemann surfaces wrapped around two EPs (red circles).
    The spectrum is purely real on the $(x,y)$-plane, and, therefore, on the corresponding $4$D hyperboloid in the system parameter space described in \eqref{sys_param}.}
    \label{fig2}
\end{figure}

The Hamiltonian in \eqref{H} does not possess a global discrete symmetry that can define its pseudo-Hermiticity. However, the pseudo-Hermitian symmetry can be expressed locally via a  Hermitian parameter-dependent matrix $\xi$, such that $ H^{\dagger}=\xi^{-1} H \xi$.
Obviously, one can take $\xi = T^{\dagger}T$, where $T$ is the diagonalizing matrix of $H$, whose columns are formed by the right eigenvector $|\psi_{1,2}\rangle$ in \eqref{psi}.

The spectrum of the NHH is characterized by the Riemann topology, where two real-valued energy surfaces wrapped around two EPs on the $(x,y)$-plane at $\vec{r}_{\rm EP}=(0,\pm 1)$ (see \figref{fig2}). Accordingly, in the system parameter space, these EPs are defined at $\epsilon_{\rm EP}=\mp 1$, $\Delta_{\rm EP}=k_{\rm EP}=0$, and $\kappa_{\rm EP}=1$. Moreover, the branch cut corresponds to the finite diabolic zero-energy line, with these two EPs on its ends (\figref{fig2}).
The right eigenvectors
become equivalent at the EPs, up to a certain global phase, namely, $|\psi_1\rangle_{\rm EP}\equiv\exp\left(\pm i{\pi}/{2}\right)|\psi_2\rangle_{\rm EP}$, respectively.

\begin{figure}[t!]
   \includegraphics[width=0.45\textwidth]{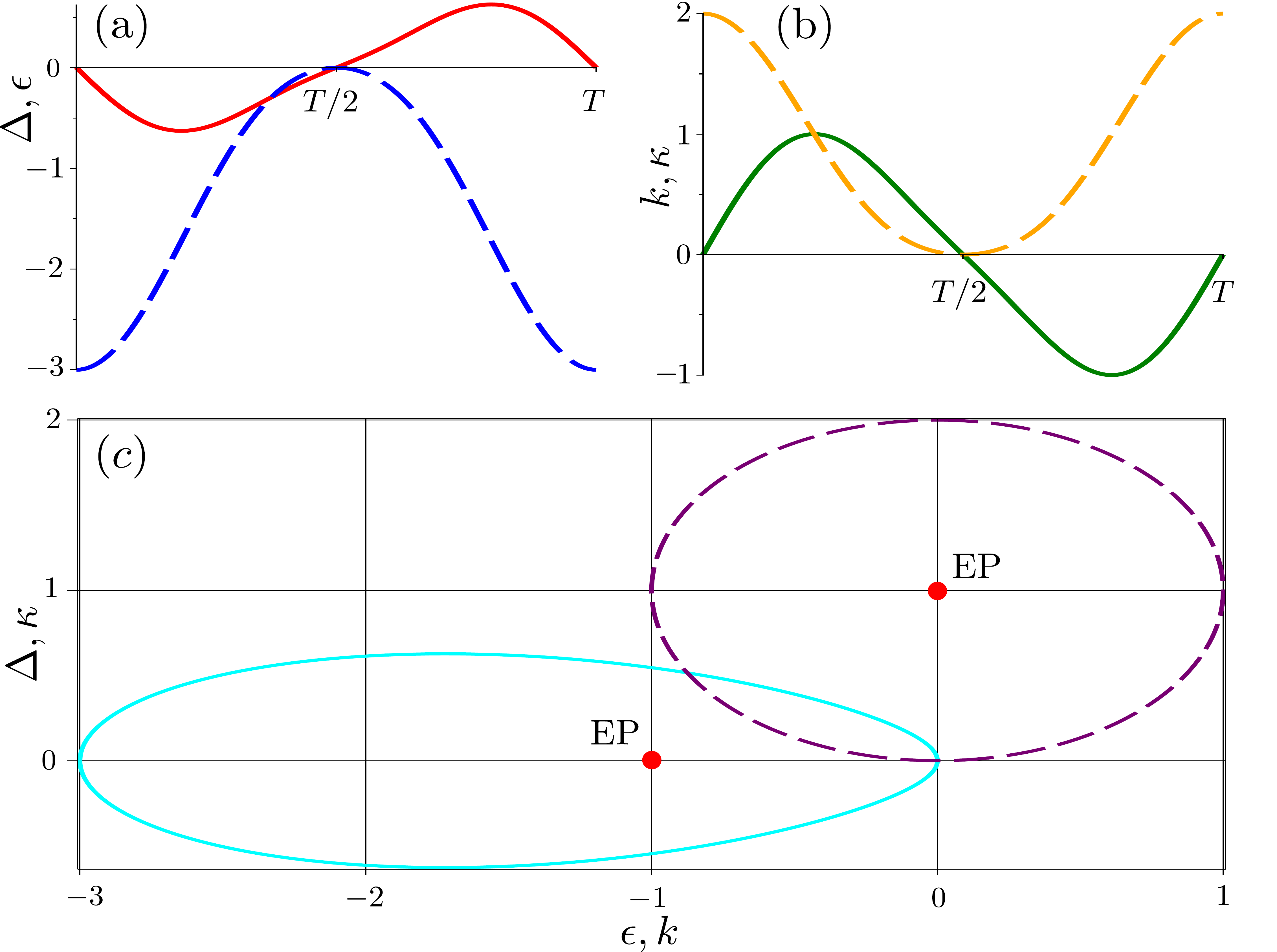}
    \caption{Time-modulated system parameters in \eqref{sys_param} when  winding in the counterclockwise direction in the chart $(x,y)$, according to \eqref{traj}. (a) Gain-loss rates $\Delta$ (red solid curve), and frequency detuning $\epsilon$ (blue dashed curve). (b) Coherent $k$ (green solid curve) and incoherent $\kappa$ (orange dashed curve) mode-coupling strengths, respectively. (c) A projection of the corresponding loop on a surface of the $4$D hyperboloid, in the system parameter space, onto the subspaces $(\epsilon,\Delta)$ (cyan solid curve) and $(k,\kappa)$ (purple dashed curve), respectively. Red points denote the same EP in both subspaces, which correspond to that in \figref{fig2} for $(x=0,y=1)$. In all panels, the winding radius is set at $r=0.5$ and the angular frequency is $\omega=2\pi$.}
    \label{fig3}
\end{figure}

\begin{figure*}
    \includegraphics[width=0.99\textwidth]{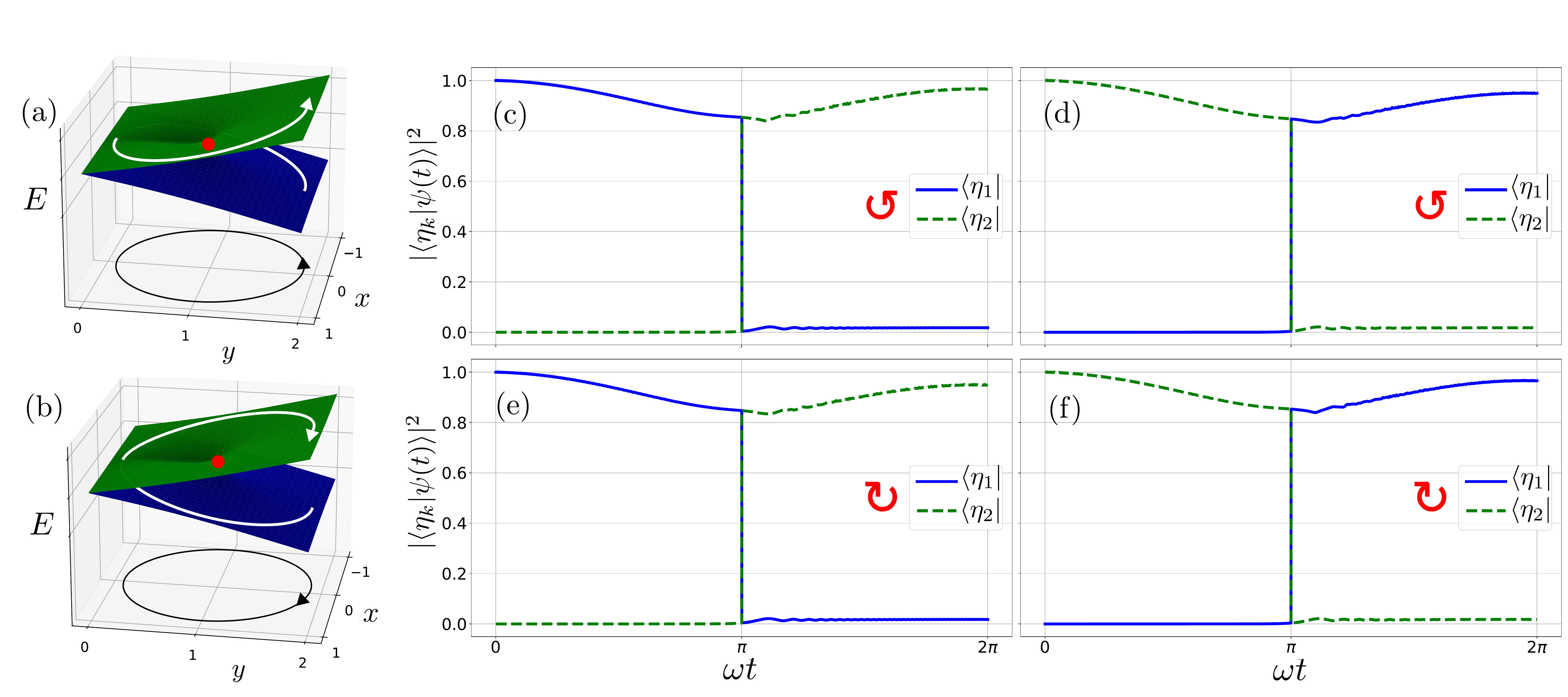}
\caption{Adiabatic state transfer while dynamically encircling around an EP in the $(x,y)$-plane. Schematic representation of the clockwise (a) and counterclockwise (b) winding direction. In (a,b) the initial state is $|\psi_1\rangle$ corresponding to the $E_1$ energy surface (blue surface). (c)-(f) Fidelity $F_{jk}=|\langle\eta_j|\psi(t)_k\rangle|^2$ between the time-evolving right eigenstate $|\psi_k(t)\rangle$ and static left eigenvector $|\eta_j\rangle$ of the NHH in \eqref{H} while encircling the EP. Panels (c)-(d) show counterclockwise winding and (e)-(f) show clockwise winding. The other parameters are: $r=0.5$, $\omega=\pi/100$, and $\phi_0=\pi$, according to \eqref{traj}. The right eigenstates are exchanged after the complete dynamical cycle regardless of the winding direction. The state dynamics exhibits a purely adiabatic character with no NATs; thus, enabling one to implement a symmetric state switch.}
    \label{fig4}
\end{figure*}

\paragraph{Adiabatic state transfer while dynamically encircling an EP.---}
Here we demonstrate that the dynamics governed by the time-dependent NHH in \eqref{H} is free from NATs, implying  
 that the states can adiabatically evolve along the orbits in the parameter space while dynamically encircling the EP. 
Let us first define the system time-evolution trajectory as:
\begin{eqnarray}\label{traj}
    x(t) &=& r\sin(\omega t+\phi_0), \nonumber \\
    y(t) &=& 1-r\cos(\omega t+\phi_0),
\end{eqnarray}
where $r,\omega,\phi_0\in{\mathbb R}$ are constants, and the time $t$ is presented in arbitrary units. Correspondingly, we change $H(t)$ in \eqref{H}.
The path in \eqref{traj} describes a circle with a radius $r$ on the plane $(x,y)$, whose center is at the EP, $\vec r_{\rm EP}=(0,1)$ [see Figs. \ref{fig4}(a,b)].  The starting point corresponds to the phase $\phi_0=\pi$, where the two energy levels $E_{1,2}$ are maximally separated. When the angular frequency $\omega>0$ $(\omega<0)$, the orbiting trajectory goes counterclockwise (clockwise). This circle trajectory on the chart corresponds to a loop on the surface of the $4$D hyperboloid  due to the embedding nature of the map $f$ (see also \figref{fig3}.)

We initialize the system in one of the right eigenstates in \eqref{psi}, namely, $|\psi(t=0)\rangle = |\psi_k\rangle$, and then we find the evolving state $|\psi(t)\rangle$ by numerically integrating the Schr\"odinger equation.
To track the state dynamics, we calculate the fidelity $F_{jk}=|\langle\eta_j|\psi(t)\rangle|^2$, i.e., the overlap of  $|\psi(t)\rangle$
with the instantaneous left eigenvector $|\eta_j\rangle$ of the NHH. 
Instantaneous in the sense that for each time step we independently calculate the left eigenvectors of the NHH in \eqref{psi_left} by substituting the parameters obtained by $(x(t), y(t))$ in \eqref{psi_left}~\cite{Zhong2018}. The results of these calculations for different winding directions and initialized states are shown in \figref{fig4}. 

As one can see in \figref{fig4}, the eigenstates $|\psi_1\rangle\to|\psi_2\rangle$ and $|\psi_2\rangle\to|\psi_1\rangle$ are exchanged after one period $T=2\pi/\omega$, regardless of the encircling direction, exhibiting thus the adiabatic character of the state evolution.
For instance, the system initialized in the state $|\psi_1\rangle$ [blue lower energy surface in Figs.~\ref{fig4}(a,b)], continuously evolves until the branch cut, corresponding to the half of the period $T/2$ [see Figs.~\ref{fig4}(c,e)]. After that, the evolving state starts moving on the upper green energy surface, corresponding to the eigenstate with energy $E_2$. Thus, after completing the full cycle, independently on the encircling path, the initial eigenstate $|\psi_1\rangle$ is switched to the eigenstate $|\psi_2\rangle$, and vice versa [see Figs.~\ref{fig4}(d,f)].
 
Such an adiabatic behavior is in striking contrast to the previously studied non-Hermitian systems with EPs, where the presence of complex Riemann topology always leads to the non-adiabatic jumps during the state evolution. Here, on the other hand, thanks to the purely real spectrum of the NHH along the chosen trajectory, the presence of NATs is eliminated, allowing to restore the adiabatic character of the non-Hermitian state dynamics.

\paragraph{Discussion.---}
Let us now comment on the stability of the observed adiabatic dynamics upon a perturbation of the NHH $H' = H + H_\delta$.
Evidently, a perturbation that breaks the pseudo-Hermiticity of the NHH in \eqref{H} may drastically affect the state evolution.
If the perturbation just shifts the NHH spectrum by an imaginary constant ($H_\delta=-i\gamma I_{2}$, where $I_2$ is the identity matrix), the adiabaticity is preserved and no NATs are induced in the system (we also confirmed this numerically). This property can be utilized, e.g., for implementing a symmetric mode converter in purely dissipative quantum systems, i.e., with no gain~\cite{Arkhipov2020,Minganti2019}.
However, by perturbing the system e.g., as  $H_\delta= {\rm diag}[\delta,0]$, where $\rm diag$ stands for a diagonal matrix, the induced difference $\nu\sim {\rm Im}\Big[\sqrt{\delta^2+4\alpha^2+4\alpha \delta\cos\phi}\Big]$ in the imaginary parts for the two eigenvalues of $H'$ becomes larger for larger $|\delta|$. The latter can eventually lead to the emergence of NATs in the system dynamics and, therefore, to a chiral transfer for $\nu\gg0$.
Furthermore, this perturbation-induced chirality, if controlled, can also enable one to switch between symmetric and asymmetric regimes on demand.

\paragraph{Conclusions.---}
We have demonstrated that time-modulated non-Hermitian systems can exhibit pure adiabatic dynamics while dynamically encircling EP in their parameter space. This is in striking contrast to previous works, where the system complex spectrum always leads to NATs during state evolution, and therefore to the asymmetric state transfer.
Remarkably, the adiabaticity can be eventually restored by properly mapping the system parameter space onto a certain manifold, over which a given NHH becomes pseudo-Hermitian with a real spectrum. 
In particular, this procedure allows to realize a long-sought symmetric mode converter, where system eigenstates are always dynamically swapped regardless of the EP winding direction. 

Evidently, the presented results also echo the adiabatic rapid passage (ARP) protocol in Hermitian systems, where a symmetric state switch is realized by adiabatically driving a system along closed loops through diabolic points (DPs) in a system parameter space~\cite{Bergmann1998,Yatsenko2002}. 
In that respect, our findings can be treated as a non-Hermitian extension of the ARP.
Indeed, the ARP protocol contains crossings of DPs, our switch protocol involves crossings of the diabolic lines while encircling the EPs, and both protocols are purely adiabatic, contrary to that in Ref.~\cite{Feilhauer2020}. 
However, the ARP  requires a series of synchronous driving optical pulses for its implementation. In contrast to this, our protocol enables the modulation of the system parameters in a single sequence, thus,  offering greater flexibility and control.
Moreover, the ARP tends to falter in the presence of decoherence. Here, in contrast, incoherent interactions play a pivotal role in achieving symmetric state conversion. 
Importantly, despite being dissipative, the spectrum of the system studied is purely real, offering greater stability compared to other NHHs with a complex spectrum.
The latter holds promising practical applications in real-world optical setups.
Our findings, thus, open new avenues for the development of novel light manipulation protocols in both classical and quantum photonics.

\acknowledgements
I.A. acknowledges support from Air Force Office of Scientific Research (AFOSR) Award No. FA8655-24-1-7376, and 
from the Ministry of Education, Youth and Sports of the Czech Republic Grant OP JAC No. CZ.02.01.01/00/22\_008/0004596.
A.M. is supported by the Polish National Science Centre (NCN)
under the Maestro Grant No. DEC-2019/34/A/ST2/00081. S.K.O.
acknowledges support from 
AFOSR Multidisciplinary University Research Initiative (MURI)
Award on Programmable systems with non-Hermitian quantum dynamics
(Award No. FA9550-21-1-0202).
F.N. is supported in part by:
Nippon Telegraph and Telephone Corporation (NTT) Research,
the Japan Science and Technology Agency (JST)
[via the Quantum Leap Flagship Program (Q-LEAP), and the Moonshot R\&D Grant Number JPMJMS2061],
the Asian Office of Aerospace Research and Development (AOARD) (via Grant No. FA2386-20-1-4069),
and the Office of Naval Research (ONR) (via Grant No. N62909-23-1-2074).


%

\end{document}